\documentclass[%
twocolumn, pra, aps, superscriptaddress, longbibliography,showpacs,amsmath,amssymb,floatfix
]{revtex4-1}

\usepackage[colorlinks,
            linkcolor=blue,
            anchorcolor=blue,
            citecolor=blue]{hyperref}
\usepackage{graphicx}
\usepackage{dcolumn}
\usepackage{xcolor}
\usepackage{multirow}
\usepackage[switch]{lineno}
\usepackage{bm}
\usepackage{physics}
\usepackage{mhchem}

\DeclareUnicodeCharacter{2212}{-}

\begin{document}
\title{Theory of vibrational Stark effect for adsorbates and diatomic molecules}
\author{Sang Yang}
\affiliation{CAS Key Laboratory of Quantum Information, University of Science and Technology of China, Hefei, 230026, China}
\author{Jun Cai}
\affiliation{Hefei National Research Center for Physical Sciences at Microscale}
\author{Yanxia Chen}
\email{yxchen@ustc.edu.cn}
\affiliation{Hefei National Research Center for Physical Sciences at Microscale}
\affiliation{Department of Chemistry, University of Science and Technology of China, Hefei, 230026, China}
\author{Ming Gong}
\email{gongm@ustc.edu.cn}
\affiliation{CAS Key Laboratory of Quantum Information, University of Science and Technology of China, Hefei, 230026, China}
\affiliation{Synergetic Innovation Center of Quantum Information and Quantum Physics, University of Science and Technology of China, Hefei, Anhui 230026, China}
\affiliation{Hefei National Laboratory, University of Science and Technology of China, Hefei 230088, China}

\begin{abstract}

Nowadays the vibrational Stark effect (VSE) of adsorbates at the electrochemical interfaces is generally investigated using the Lambert theory, in which the strong electric field across the interfaces can be treated as some kind of perturbation. Lambert found that the VSE arises mainly from the classical effect, and the quantum effect is negligible. This idea is accepted by almost all current first-principle calculations for this issue. Here we revisit this problem by addressing the fundamental question that to what extent the quantum effect is important for VSE, and if it is observable, then which physical quantity determines this effect. We use the Morse, Lennard-Jones and Dunham potentials as basic potentials to explore this problem using quantum perturbation theory. We define the relative difference between quantum and classical VSE slopes to define the quantum effect, $\eta$, and show that for CO, $\eta \sim $ 2 - 3\%, while for adsorbed hydrogen on Pt electrode, $\eta \sim$ 8 - 10\%, using the experimental data. We find that $\eta$ is determined by the anharmonic coefficient $\chi_e$. Without results we present a new understanding of the VSE as a function of electric field and potential in electrochemical experiments, showing that the nonlinear slope of VSE as a function of potential should arise from the nonlinear relation between electric field and potential across the interfaces, which may resolve the long-standing controversial in experiments. 
\end{abstract}

\maketitle

\section{Introduction} 
\label{sec-introduction} 

\begin{figure}[htp]
    \centering
	\includegraphics[width=0.35\textwidth]{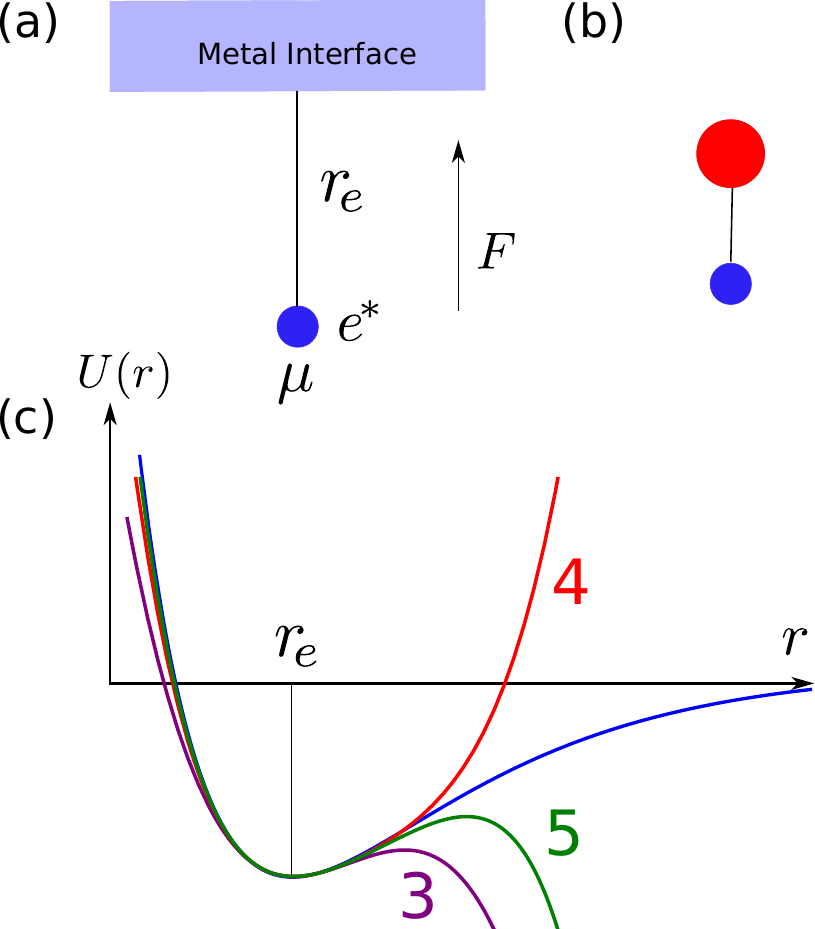}
	\caption{(a) Atoms suspended at the metal-solution interface with reduced mass $\mu$ and effective charge $e^{\ast}$ (thus $M_1 = -e^* r$ in Lambert model). The interfacial electric field $F$ is perpendicular to the interface. (b) The diatomic molecule in an electric field. (c) Morse potential and the 3rd (purple), 4th (red) and 5th (green) order truncation of the Dunham potential versus $r$, with equilibrium position $r_e$.}
    \label{fig-fig1}
\end{figure}

When an atom \cite{toya1962infrared,jayasooriya1980vibrational,xu1999raman}, or a molecule \cite{hoffmann1983infrared} is adsorbed on an electrode surface (Fig. \ref{fig-fig1}(a)), its vibration can provide invaluable information for us to understand the interaction \cite{mahan1978collective, persson1979absorption, persson1980collective, persson1981collective, persson1981vibrational} between the adsorbate and electrode surfaces \cite{Holloway1984changes, Weaver1993electrostatic, Wasileski2002field}. This is also one of the simplest physical models that can be conceived in theory, which has found important applications in electrocatalysis \cite{Kunimatsu2006situ, Kunimatsu2007hydrogen, Zhu2020pHdependent, Zhu2022Electrolyte}. The vibrational spectra as well as its potential dependence may reveal rich information about reaction intermediates involved in the electrocatalytic reactions \cite{nichols1988spectroscopic, Kunimatsu2006situ, Kunimatsu2007hydrogen}. The theoretical understanding of potential dependence of the vibrational frequency, i.e., the electrochemical Stark effect, can be traced back to 1983 when David Lambert pioneered a novel experimental method, the electroreflectance vibrational spectroscopy (EVS) method, to observe the vibration of carbon monoxide (CO) at the metal-CO interface \cite{Lambert1983Observation}. Prior to this work, there were already some experiments investigating the vibrational behavior of CO on metal surfaces \cite{Erley1979CO, Ortega1982adsorption}, however, most of these early works were predominantly within the realm of theoretical physics \cite{ray1982variations, toya1962infrared, rose1981universal, rose1983universal}. Lambert was the first to systematically study the vibrational spectra with a controlled interfacial electric field strength based on a nonlinear classical method \cite{Lambert1983Observation, Lambert1984Stark,Lambert1985co, Lambert1988vibrational,Lambert1990co, Wang1994coadsorption, Luo1990co, Luo1993electric,lambert1991electric,Lambert1996Vibrational}, in which the 
vibrational Stark effect (VSE) as a function of nonlinearity parameters is explicitly obtained. It has been accepted for a long time that this method is sufficient to understand the VSE since the quantum contribution is negligible. This is also the theoretical model adopted in first-principle calculation \cite{Hamada2008Density, koper2011combining, weaver1999binding}, where many nonlinearity parameters are used to explain the VSE. Unfortunately, these high-order nonlinear parameters can not be directly determined in experiments. Furthermore, this treatment also introduces some form of dilemma in the theoretical calculations. On one hand, we have to use the full quantum theory to understand the physics in this model; yet on the other hand, we use the classical theory to determine the VSE, assuming that the quantum effect is negligible.

The Similar issue of VSE also exists in diatomic molecules (Fig. \ref{fig-fig1}(b)) \cite{pliskin1960infrared, walsh1951ii} in free space, in which the vibrational spectra and the electric field can be measured or tuned with much higher accuracy. Meanwhile, it can be much easier studied using the theory of quantum mechanics. According to this theory, the vibrational frequency is well-known to be
\begin{equation}
\omega_n = (n+{1\over 2})\omega - \frac{1}{4}\chi_e (n+{1\over 2})^2 \omega, 
\end{equation}
where $\omega$ is the vibrational frequency, and $\chi_e$ is the dimensionless anharmonic coefficient from the nonlinear effect. This formula may be traced back to 1929 in the Morse potential \cite{Morse1929Diatomic}, in which $\chi_e$ may take a simple form from dimension analysis; see below. Thus these two different systems, which share great similarities, are studied using different approaches, and to date, to what extent the quantum effect is important in these two systems is still unresolved. For example, the parameter $\chi_e$ will not enter the Lambert classical theory of VSE due to the absence of the Planck constant. 

This work aims to revisit this problem, focusing on the role of the quantum effect on the vibrational frequency and its Stark effect. For adsorbed molecules, the physics is much more complicated because the nonlinearity may be influenced by a lot of factors, which are totally undeterminable in experiments. Thus, to obtain some conclusive conclusions, the simplest way to address this issue is to study the quantum effect at various potentials. We define the quantum effect $\eta$ by the relative difference between quantum and classical VSE slopes. Our major results are summarized as follows. 
(I) For the hydrogen atom, the quantum effect can reach about 8 - 10\%, while for CO, its effect can be around 2 - 3\%, using the parameters available in experiments. The Lambert theory is insufficient to capture this effect due to the absence of the fifth-order term in the polynomial expansion of the potential. 
(II) The quantum effect $\eta$, which is a dimensionless constant, is comparable with the anharmonic coefficient $\chi_e = \hbar\omega/D_e$, where $\hbar$ is the Planck constant divided by $2\pi$, $\omega$ is the vibrational frequency and $D_e$ is the binding energy. We examine various molecules and find that $\eta \sim \chi_e$. Our results clarify the quantum effect in VSE and establish a clear relation between anharmonic coefficient and quantum effect in VSE. (III) Our results also mean that in experiments the nonlinear Stark effect as a function of potential, as observed in electrochemical vibrational spectra \cite{zhang2016nonlinear,staffa2017determination,tong2023linear}, should come from the nonlinear relation between the electric field and potential across the metal-solvent interface.

This manuscript is organized as follows. We present a review of the Lambert theory in Sec. \ref{sec-reviewlamberttheory}, and then we present our model based on quantum mechanics in Sec. \ref{sec-model} and our
calculated results in Sec. \ref{sec-calculation}. Three remarks are presented in Sec. \ref{sec-remarks}, followed by a summary and discussion of the results in Sec. \ref{sec-conclusion}. A detailed analysis of $\eta$ from different approximations is presented in Appendix \ref{sec-appendix}. 

\section{Review of the Lambert theory}
\label{sec-reviewlamberttheory}

The VSE can be effectively modeled using an electrostatic interaction with a strong interfacial electric field \cite{helmholtz1853ueber, guoy1910constitution, chapman1913li}, and the corresponding effective Hamiltonian \cite{Lambert1983Observation} can be 
written as 
\begin{equation}
H=\frac{p^2}{2\mu}+U(r)-M(r)F.
\end{equation}
Here $p$ is the momentum, $\mu$ is the reduced mass of the diatomic molecule, $U(r)$ is the potential (see Eq. \ref{eq-UrMr}), and $M(r)$ is the dipole moment. Focusing on the stretching vibration along the normal direction, the potential and dipole moment can be expanded as a power series of $r$, retaining only the leading terms. The potentials take the forms \cite{Lambert1983Observation, Lambert1984Stark}
\begin{eqnarray}
U(r)&=&U_0+{1\over 2} K_2 (r-r_e)^2+K_3 (r-r_e)^3+\ldots \\
\label{Ur_duhelm}
M(r)&=&M_0+M_1 (r-r_e)+M_2 (r-r_e)^2+\ldots.
\label{eq-UrMr}
\end{eqnarray}
Here, several parameters $K_2$, $K_3$, $M_1$ and $M_2$ are introduced, and $F$ is the electric field strength at the interface. In this potential, the quadratic term $K_4(r-r_e)^4$ is neglected, assuming it to be very small. The interfacial electric field induces a small change of the equilibrium position from $r_e$ to $r_e'$, subsequently altering the corresponding vibration frequency. The above model can be solved 
by determining the equilibrium position at 
\begin{equation}
r_e' = {-K_2 + 2F M_2 + \sqrt{12 F K_3 M_1 + (K_2 - 2F M_2)^2} \over 6K_3},
\end{equation}
with $r_e'(F = 0) = 0$. The vibration frequency is given by $\mu {\omega}^2/2 = \sqrt{3F K_3 M_1 + (K_2/2 - F M_2)^2}$. To the leading-order term, one can have
\begin{equation}
\omega = \sqrt{K_2 \over \mu} + {F (3 K_3 M_1 - K_2 M_2) \over 
K_2^{3/2} \sqrt{\mu}} + \mathcal{O}(F^2),
\end{equation}
and the key formula is the slope of the VSE, which is given by
\begin{equation}
\frac{d\omega}{dF} = \omega {(3K_3 M_1  - K_2 M_2 ) \over K_2^2}.
\label{Lambert_methods_result}
\end{equation}
This result means if $K_3 = 0$ and $M_2 = 0$, or if $M_1=0$ and $M_2 = 0$, the slope is zero. Furthermore, the slope is nonzero if $K_3 = 0$ yet the nonlinear effect in the dipole momentum is nonzero. The minus sign here implies the opposite role played by the nonlinearity of the potential and the dipole momentum. 

This model has been widely used in first-principle simulations \cite{wasileski2001metal, Kunimatsu2007hydrogen, Hamada2008Density}. Using the parameters in \cite{Lambert1984Stark}, one finds that $K_3$ is the leading contribution to the VSE. At the interface, the electric field strength can reach $F \sim 10^9$ V/m \cite{Lambert1984Stark,Weaver1993electrostatic,Wasileski2002field,Kunimatsu2006situ,Kunimatsu2007hydrogen}, in which the nonlinear VSE is given by 
\begin{equation} 
\frac{d^2 \omega}{d F^2}=\omega\frac{18K_2K_3M_1M_2-27K_3^2M_1^2-K_2^2M_2^2}{2K_2^4}.
\label{eq-secondorderdw2}
\end{equation}
Based on the perturbation theory, we can write
\begin{equation}
\omega(F) = \omega_0 + \delta \omega + \delta \omega^{(2)} +\cdots,
\end{equation}
then the ratio between the quadratic shift $\delta \omega^{(2)} = {1\over 2 }\frac{d^2 \omega}{d F^2} F^2$ 
and the linear Stark shift $\delta \omega  = \frac{d \omega}{d F}F$ is estimated to be $\delta \omega^{(2)}/\delta \omega \sim 10^{-3}$ using the experimental available parameters. Thus the linear Stark shift is always dominated. It is unlikely that changing the parameters can induce significant nonlinear Stark effect. A further estimations shows that when the nonlinear Stark effect is important, the electric field $F$ should be comparable to its atomic unit $F_0 = 5.14\times 10^{11}$ V/m. For this reason, the following discussion only considers the first-order effect.

\section{Our Theoretical Model}
\label{sec-model}

\begin{figure}[htp]
\centering
\includegraphics[width=0.35\textwidth]{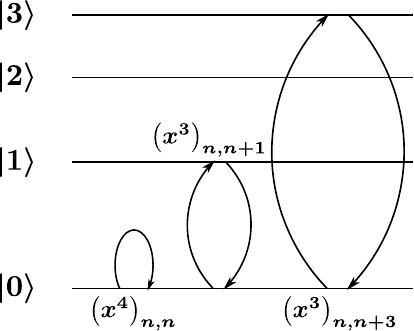}
\caption{Contribution of terms $x^3$, $x^4$ to VSE via perturbation theory. }
\label{fig-fig2}
\end{figure}

Our idea is to examine the validity of the above theory using quantum mechanics. To this end, let us first consider the solution of the following model with the Dunham potential (the 3rd, 4th and 5th order truncations of this potential are shown as Fig. \ref{fig-fig1}(c))
\cite{Dunham1932EnergyLevel}. The divergence of potential in the large $r$ limit is a common feature in all polynomial expansion approximation \cite{simons1973new}; it may be avoided using Pade approximation. We have 
\begin{equation}
H = {p^2 \over 2\mu} +U_\text{D}, \quad U_\text{D} = {K_2 \over 2} x^2 + K_3 x^3 + K_4 x^4 + K_5 x^5, 
\end{equation}
where $x=r-r_e$. Different from the Lambert model, here we have taken $K_4$ into account. The reason is that, while $K_3$ term is much more important, it contributes to the spectra from a second-order transition process. However, $K_4$ term contributes to the spectra from the first-order process; see the transition diagram in 
Fig. \ref{fig-fig2}. This result is well-known in nonlinear vibration physics \cite{Landau1976Mechanics}, which has not been taken into account in the Lambert theory. We find 
\begin{eqnarray}
E_n &=& \hbar \omega_0(n+\frac{1}{2}) - \frac{15}{4} \frac{K_3^2}{\hbar \omega_0}(\frac{\hbar}{\mu \omega_0})^3(n^2+n+\frac{11}{30})  \nonumber \\
&&+ \frac{3}{2} K_4(\frac{\hbar}{\mu \omega_0})^2(n^2+n+\frac{1}{2}),
\label{quantum_nenergy_level}
\label{energy-En}
\end{eqnarray}
where the classical frequency $\omega_0 = \sqrt{K_2/\mu}$. The $K_5$ term contributes to the system via a second-order process, which is too small to be considered. However, it will contribute to the VSE. 

\begin{table*}[htp]
\label{table-VSE}
	\centering
	\caption{Summary of force constants in the nonlinear potentials $U_\text{D}$, $U_\text{Morse}$, $U_\text{LJ}$ and $U_\text{Mie}$ potentials, with equibirum position at $r_e$ at $F = 0$. In the Dunham potential, we have presented the modification of the parameters by $K_5$. }
	\begin{tabular}
		{|p{0.05\textwidth}|p{0.17\textwidth}|p{0.15\textwidth}|p{0.25\textwidth}|p{0.29\textwidth}|} \hline
	    &    $U_{\text{Morse}}$ & $U_{\rm{D}}$ & $U_{\rm{LJ}}$ &$U_{\rm{Mie}}$ \\ \hline
		$r_e^{\prime}$  & $r_e- \frac{e^{\ast}}{2 a^2 D_e}F$  & $r_e - {e^*\over K_2}F$    & $r_e-\frac{e^{\ast} r_e^2}{2n^2 D_e}F$ &$r_e-{e^*r_e^2\over D_e A n^2}F$  \\ \hline
		$K_2^{\prime}$  & $2 a^2 D_e+3 a e^{\ast} F$  & $K_2 - {6e^* K_3 \over K_2}F$ & $\frac{2n^2 D_e}{r_e^2}+\frac{3(n^2+n^3) e^{\ast}}{n^2r_e}F$ &$\frac{AD_en^2}{r_e^2}+\frac{(3+An+n)e^*}{r_e}F$ \\ \hline
		$K_3^{\prime}$  & $- a^3 D_e-\frac{7}{6}a^2e^{\ast}F$  & $K_3 - {4e^* K_4 \over K_2}F$  & \begin{tabular}[c]{@{}l@{}}$-\frac{(n^2+n^3)D_e}{r_e^3}$\\$-\frac{(11n^2+18n^3+7n^4) e^{\ast}}{6n^2 r_e^2}F$\end{tabular} &\begin{tabular}[c]{@{}l@{}}$-\frac{AD_en^2(3+An+n)}{6r_e^2}$\\ $+\frac{(11+n(6+n+A(6+n+An)))e^*}{6r_e^2}F$\end{tabular} \\ \hline
		$K_4^{\prime}$  & $\frac{7}{12}a^4D_e+\frac{5}{8}a^3e^{\ast}F$  & $K_4 - {5e^* K_5 \over K_2}F$  & \begin{tabular}[c]{@{}l@{}}$\frac{(11n^2+18n^3+7n^4) D_e}{12r_e^4}$\\ $+\frac{5(10n^2+21n^3+14n^4+3n^5) e^{\ast}}{24n^2 r_e^3}F$\end{tabular} &\begin{tabular}[c]{@{}l@{}}$\frac{D_en (A (A n+n+6)+n+6)+11}{24r_e^4}$\\ $+\frac{(5+n+An)(10+n(5+n+A(5+An)))e^*}{24r_e^3}F$\end{tabular} \\ \hline
	\end{tabular}
 \label{table-vse}
\end{table*}

To examine the validity of the above approach, we next consider the physics in the Morse potential and the LJ potential, in which the parameters have been well determined in experiments, thus the quantum effect can be seen much clearer. These two models have also been widely studied in supersymmetry \cite{Witten1982, Woods1954Diffuse, Eckart1030Penetration,poschl1933bemerkungen, rosen1932vibrations, ahmadov2018analytical, Ample2002Adsorption}. The Hamiltonian of this oscillator under a strong interfacial electric field $F$ can be expressed in the following form (set $M_1 = -e^* r$)
\begin{equation}
	H=\frac{p^2}{2\mu}+U(r)+e^{\ast}Fr.
        \label{ham_elec}
\end{equation}
We consider the following potentials, which have been widely used in literature \cite{Morse1929Diatomic, Lambert1984Stark, Lambert1983Observation,korzeniewski1986theoretical}
\begin{eqnarray}
U_\text{Morse}(r)&=&D_e[1-e^{-a(r-r_e)}]^2,\label{potential-morse}, \\ 
U_{\text{LJ}}(r)&=&D_e((\frac{r_e}{r})^{2n}-2(\frac{r_e}{r})^{n}), \\
U_{\text{Mie}}(r)&=&\frac{D_e}{A-1}((\frac{r_e}{r})^{An}-A(\frac{r_e}{r})^{n}).
\label{potential-lj}
\end{eqnarray}
In $U_{\text{LJ}}(r)$ with $n=6$, it corresponds the 12-6 LJ potential; and the Mie potential $U_{\text{Mie}}(r)$ with two empirical parameters $A$ and $n$ 
is a direct generation of LJ potential $U_{\text{LJ}}(r)$. These potentials will be assumed to have the same binding energy, equilibrium bond length and vibrational frequency, which are directly determined by experiments. Their different nonlinear terms may lead to slightly different VSE in a strong electric field. 

We solve the spectra of the above models using Eq. \ref{quantum_nenergy_level}. Under such conditions, the external electric field $F$ can modify the equilibrium position slightly. We assume $r_e' = r_e + B F + \mathcal{O}(F^2)$, then we model the potential as 
\begin{equation}
U(x) = U_0 + {K_2' \over 2} x^2 + K_3' x^3 + K_4' x^4, 
\label{potential_electric}
\end{equation} 
where $x = r - r_e'$ and $U_0 = U(r_e)$. The key idea is that $K_4$ can be modified by the nonlinearity term of $K_5(r-r_e')^5$, which has been ignored in all the previous literature. For these models, the corresponding parameters of $K_i'$ and $r_e'$ are summarized in Table \ref{table-vse}. In this calculation, we find that the modification of $K_4'$ comes from the fifth-order $K_5$, which has been ignored in the Lambert theory. Using these expressions in Table \ref{table-vse}, we can determine the vibrational frequency, in which $\hbar\omega = E_1 - E_0$. 

\section{Calculate vibrational Stark effect}
\label{sec-calculation}

\begin{figure*}[htp]
\centering
\includegraphics[width=0.66\textwidth]{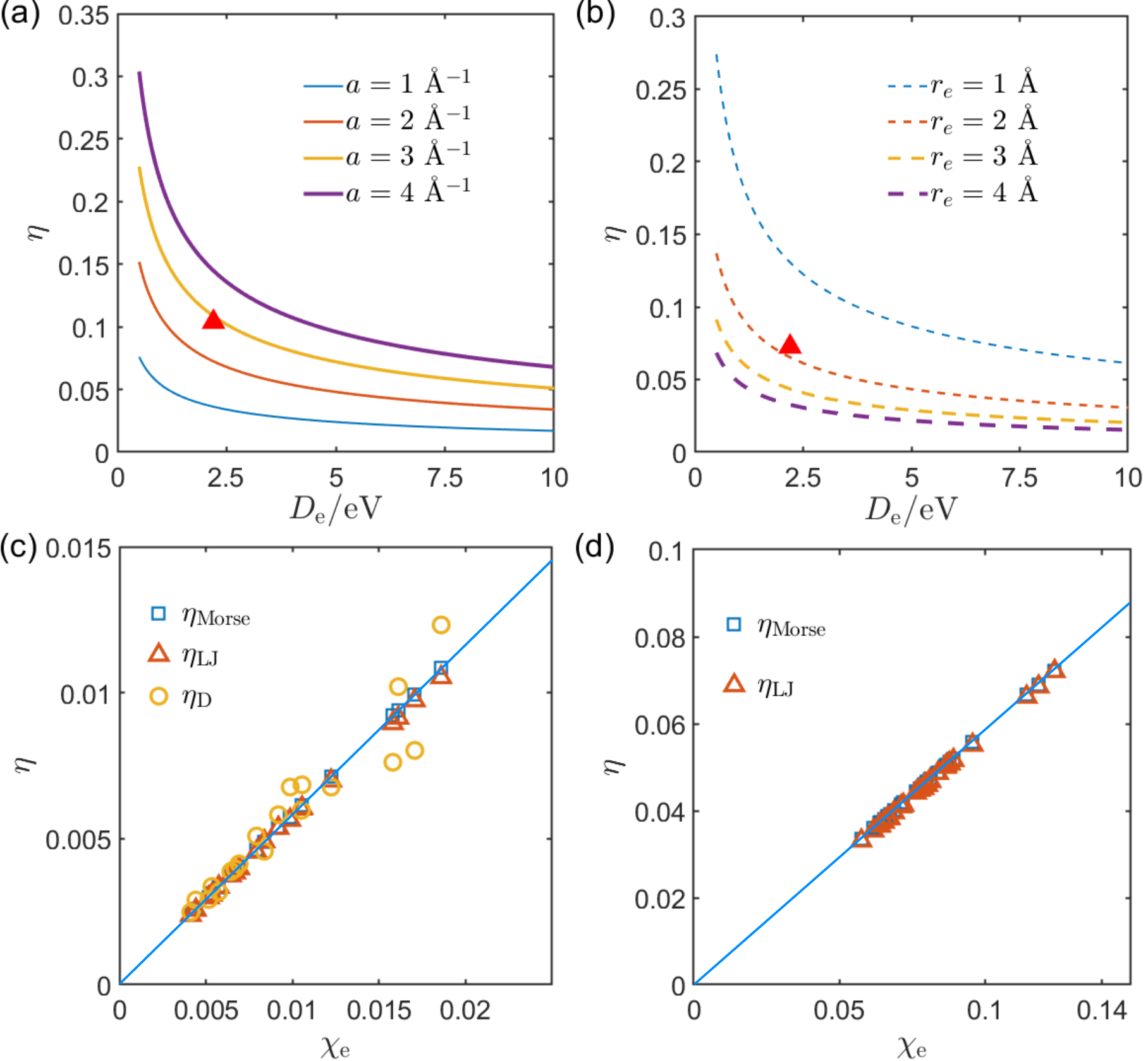}
\caption{Quantum effect $\eta$ as a function of binding energy $D_e$ for various $a$ and $r_e$. The red triangle are results for Pt-H using 
\cite{Kunimatsu2007hydrogen} $a=2$ \AA$^{-1}$, $D_e=2.28$ eV and $r_e=1.8$ \AA. (a) Results for Morse potential; and (b) results for 12-6 LJ potential. (c) and (d) are quantum effect as a function of $\chi_e$ for non-hydride diatomic molecules and hydride diatomic molecules, respectively. The data are summarized in Tables \ref{table-nonH} and \ref{table-Hmoculars}.}
\label{fig-fig3}
\end{figure*}

Using the above results we can determine the vibrational frequency analytically. Since only the linear VSE is important, we can define 
\begin{equation}
\omega(F) = \omega_0 + {d\omega \over dF} F,
\end{equation}
with
\begin{eqnarray}
\omega_0|_\text{Morse} &=& a\sqrt\frac{D_e}{\mu}-\frac{a^2\hbar}{\mu},\\
\omega_0|_\text{D} &=& \sqrt{\frac{K_2}{\mu}} - \frac{15\hbar K_3^2}{2K_2^2\mu} + \frac{3\hbar K_4}{K_2\mu}, \\ 
\omega_0|_\text{LJ} &=& \frac{n}{r_e}\sqrt{\frac{2D_e}{\mu}}-\frac{15(n^2+n^3)^2}{8n^4r_e^2}\frac{\hbar}{\mu} \nonumber\\
&&+ \frac{11n^2+18n^3+7n^4}{8n^2r_e^2}\frac{\hbar}{\mu}, \\
\omega_0|_\text{Mie} &=& \frac{n}{r_e}\sqrt{\frac{AD_e}{\mu}}+\frac{12+12(1+A)n}{24r_e^2}\frac{\hbar}{\mu}\nonumber\\
&&+ \frac{(2+A(7+2A))n^2}{24r_e^2}\frac{\hbar}{\mu},
\end{eqnarray}
and the slope of VSE are 
\begin{eqnarray}
{d\omega \over dF}|_{\text{Morse}} &=& \frac{3e^{\ast}}{2\sqrt{2}}\sqrt{\frac{1}{D_e\mu}}+\frac{7}{8}\frac{ae^{\ast}\hbar}{D_e\mu},\\
\label{eq-21}
\end{eqnarray}
\begin{eqnarray}
{d\omega \over dF}|_{\text{D}} &=& -\frac{3e^*K_3}{K_2^2}\sqrt{\frac{K_2}{\mu}}
-\frac{15\hbar e^*K_5}{K_2^2\mu}\nonumber\\
&&- \frac{3\hbar e^*(30K_3^3-26K_2K_3K_4)}{K_2^4\mu},\\
{d\omega \over dF}|_{\text{LJ}} &=& \frac{3(1+n)e^{\ast}}{2\sqrt{2}n}\sqrt{\frac{1}{D_e\mu}}\nonumber\\
&&+\frac{(1+n)(1+2n)(7n-3)}{16n^2D_er_e}{\hbar e^{*}\over\mu},\\
{d\omega \over dF}|_{\text{Mie}} &=& \frac{(3+n+An)e^{\ast}}{2n}\sqrt{\frac{1}{AD_e\mu}}\nonumber\\
&&-\frac{9+(2+2A)n}{24AD_en^2r_e}{\hbar e^{*}\over\mu}\nonumber\\
&&+\frac{15A+3A^2-(7A+7A^2)n}{24AD_er_e}{\hbar e^{*}\over\mu}.
\label{eq-24}
\end{eqnarray}
In the above expressions, the terms without $\hbar$ are results that can be described well using classical mechanics, and the terms with $\hbar$ should come from the quantum effect. Thus in these expressions, the quantum effect and classical effect are all included. With these results, let us come back to the Lambert theory \cite{Lambert1983Observation, Lambert1984Stark, Lambert1996Vibrational}, in which the nonlinear coefficients of $K_4$ and $K_4'$ are neglected. We have 
\begin{eqnarray}
{d\omega \over dF}|_{\text{Morse}} &=& \frac{3e^{\ast}}{2\sqrt{2}}\sqrt{\frac{1}{D_e\mu}}, \\
{d\omega \over dF}|_{\text{D}} &=& -\frac{3e^*K_3}{K_2^2}\sqrt{\frac{K_2}{\mu}}, \\ 
{d\omega \over dF}|_{\text{LJ}} &=& \frac{3(1+n)e^*}{2\sqrt{2}n}\sqrt{\frac{1}{D_e\mu}},\\
{d\omega \over dF}|_{\text{Mie}} &=& \frac{(3+n+An)e^*}{2n}\sqrt{\frac{1}{AD_e\mu}}.
\end{eqnarray}
These terms correspond to the results in Eq. \ref{eq-21} to Eq. \ref{eq-24} by setting $\hbar = 0$.

Then we can define the quantum effect as the relative difference between quantum slope and classical slope as 
\begin{equation}
\eta = {{d\omega \over dF}|_{\text{Quantum}} - {d\omega \over dF}|_{\text{Classical}}
\over {d\omega \over dF}|_{\text{Classical}}}.
\label{ratio_quantum_correction}
\end{equation}
By the dimensional analysis, we expect $\eta$ to be a function of $a$, $\hbar$, $D_e$ and $\mu$, and the only possible form is (see Appendix \ref{sec-appendix})
\begin{equation}
\eta \sim {\hbar\omega \over D_e} \sim \frac{a\hbar}{\sqrt{D_e\mu}}.
\end{equation}
It is well known that the anharmonic coefficient $\chi_e$ is also defined as $\hbar\omega/D_e$. Thus there is an intimate relation between the quantum effect and the anharmonicity. For the Morse potential, LJ potential and general Mie potential, we have 
\begin{eqnarray}
    \eta_\text{Morse} &=& \frac{7}{6\sqrt{2}} \frac{a\hbar}{\sqrt{D_e\mu}} = \frac{7}{12} {\hbar\omega \over D_e},\\
    \label{eta-morse}
    \eta_\text{LJ} &=& \frac{-3+n+14n^2}{12\sqrt{2}n} \frac{\hbar}{r_e\sqrt{D_e\mu}} \nonumber\\
    &=& \frac{-3+n+14n^2}{24n^2} {\hbar\omega \over D_e}, 
    \label{eta-lj}
\end{eqnarray}
\begin{eqnarray}
    \eta_{\rm{Mie}}&=&[\frac{-9+n(-2+3n)}{12(3+n+An)An^2}\nonumber\\
    &&+\frac{-2+3(5+A)n+7(1+A)n^2}{12n(3+n+An)}]\frac{\hbar\omega}{D_e}.
    \label{eta-mie}
\end{eqnarray}

\begin{figure}[htp]
\centering
\includegraphics[width=0.42\textwidth]{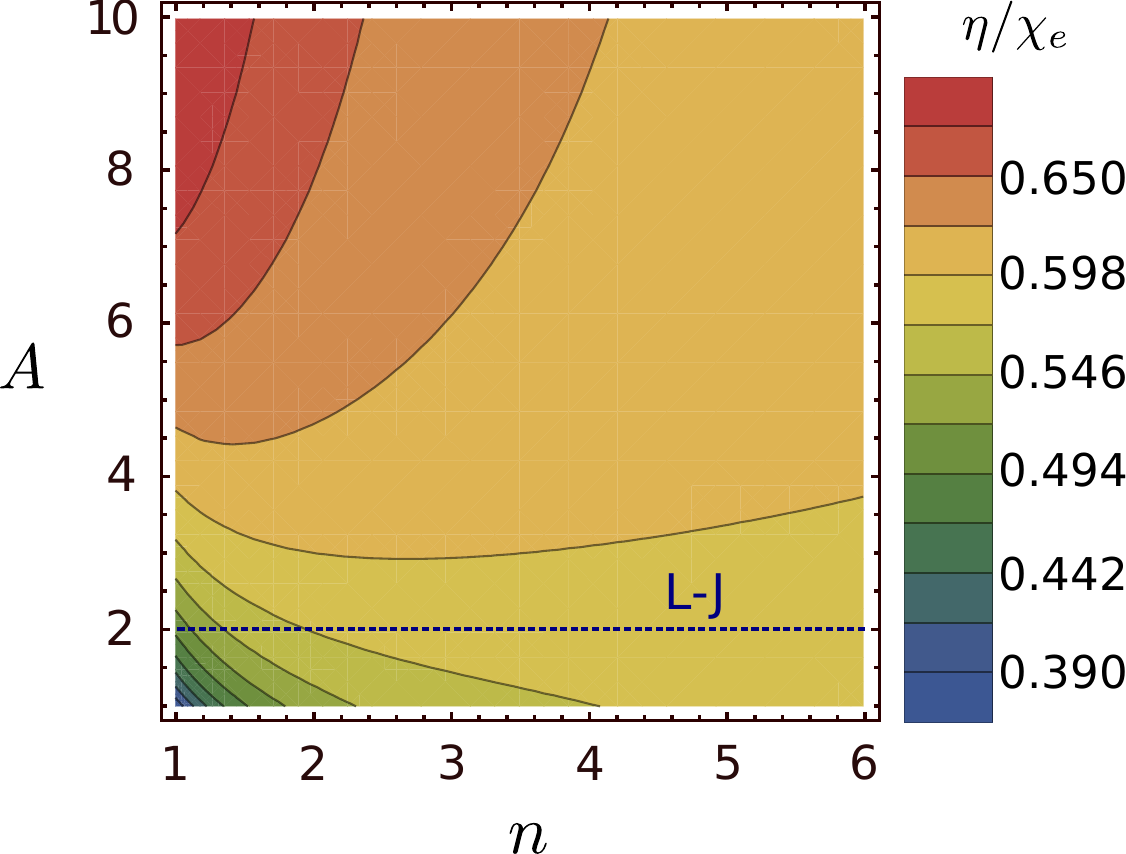}
\caption{Quantum effect in the Mie potential as a function of $n$ and $A$. The dashed line corresponds to the LJ potential, with $A=2$.}
\label{fig-fig4}
\end{figure}
The results for some chosen $a$ and $r_e$ are presented in Fig \ref{fig-fig3} (a) and (b). In these two figures, we have also given the results for Pt-H vibration with $D_e=2.28$ eV and $a=2$ \AA, which have been determined in experiments \cite{Kunimatsu2006situ, Kunimatsu2007hydrogen}, showing that $\eta \sim 10\%$. We also summarized the data of different molecules in Table. \ref{table-nonH} (without hydrogen atom) and Table. \ref{table-Hmoculars} (with the hydrogen atom, named as hydride molecule), which have been determined in experiments, and plotted the quantum effect in Fig. \ref{fig-fig3}(c) and (d), respectively. We find that the increasing of binding energy will reduce the quantum effect. Furthermore, for different atoms, $\eta \sim \chi_e$, where $\eta$ 
is determined by our theory, and $\chi_e$ is given by experiments. For comparison, we have also presented the quantum effect from different approximations in Appendix \ref{sec-appendix}. 

\begin{figure}[htp]
\label{fig-fig5}
\centering
\includegraphics[width=0.4\textwidth]{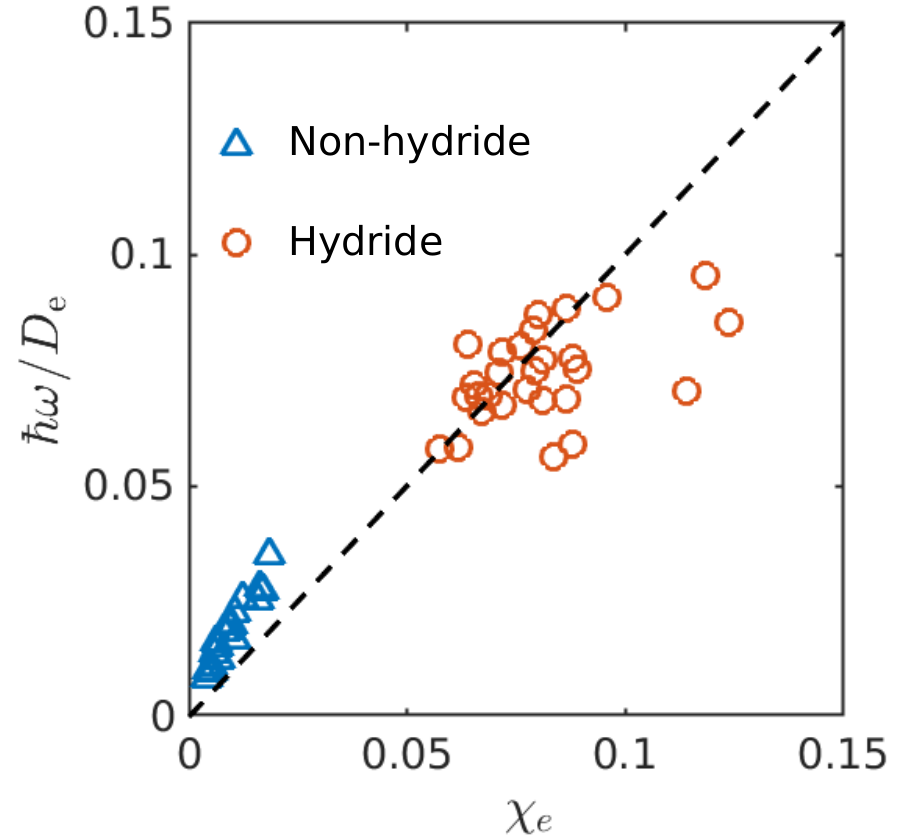}
\caption{Comparison of experimental and theoretical nonlinearity for hydride diatomic molecules and non-hydride diatomic molecules. These data are taken from Table. \ref{table-nonH} and Table. \ref{table-Hmoculars}. The dashed line gives $\hbar\omega/D_e = \chi_e$. }
\label{fig-fig5}
\end{figure}

In our determination of the parameters in the generalized LJ potential and Mie potential, we find $n \sim 2 - 3$, which is significantly smaller than the 12-6 LJ potential used in a lot of theories and experiments. The reason is that we have used the experimental data of $D_e$ and $r_e$, thus $n$ is used to determine the vibrational frequency $\omega$. Take the results for LiF as an example. It has $D_e = 5.96$ eV and $r_e = 1.563$ \AA, with vibrational frequency $\omega/2\pi c = 910.34$ cm$^{-1}$, and $\mu = \mu_H /(1/7 + 1/19) = 5.1\mu_H$ ($\mu_H = 1800 m_e$ is the mass of hydrogen atom), we have
\begin{equation}
n = {\sqrt{\mu c^2} r_e \omega \over \sqrt{2D_e} c} = 1.74.
\end{equation}
Furthermore, let the parabolic potential from LJ potential and Morse potential be the same, we have $D_e a^2 (r-r_e)^2 = D_e n^2 (r-r_e)^2/r_e^2$, thus $a = n/r_e = 1.74/1.563 = 1.11$ \AA$^{-1}$. This expression means that the index $n$ depends strongly on the $\mu$, $r_e$, $\omega$ and $D_e$ and there is no reason that it can be a universal constant. The same way can be used to determine $a$ in the Morse potential; see Table \ref{table-nonH} and Table \ref{table-Hmoculars}.

While the parameters in the LJ and Morse potential can be uniquely determined in the above way, the Mie potential contains two parameters, which can not determined simultaneously. Let the parabolic potential $\mu \omega^2 (r-r_e)^2/2 = A D_e n^2 (r-r_e)^2/(2r_e^2)$, we have 
\begin{equation}
A = {\mu r_e^2 \omega^2  \over D_e n^2} = {\mathcal{Q} \over n^2}.
\end{equation}
For LiF, we have $\mathcal{Q} = 12$; and for CsI, we have 
$\mathcal{Q} =20$. Thus in Fig. \ref{fig-fig4}, we plot $\eta/\chi_e$ as a function $n\in (1, 6)$ and $A \in (1, 10)$, showing that for different parameters, this ratio is in the range of 0.54 to 0.69. This result means that for different system parameters, we always have the conclusion that $\eta \sim \chi_e$. Since $\chi_e$ in diatomic molecules are determined in experiments from the vibrational spectra, and $\hbar\omega/D_e$ are determined by some different approaches, in Fig. \ref{fig-fig5}, we plot the relation between $\hbar\omega/D_e$ as a function of $\chi_e$, showing that $\chi_e \sim \hbar\omega/D_e$. These results 
demonstrate the generality between the quantum effect and anharmonicity for the understanding of VSE. 

\section{Three Remarks}
\label{sec-remarks}

Three remarks about the above results are presented below. Firstly, let us clarify why in the Lambert theory for CO, the quantum effect is estimated to be 0.3\% \cite{Lambert1988vibrational, Lambert1996Vibrational}, while in this work it is greatly enhanced.  The reason is that, while the contribution of nonlinear effect to the vibrational spectra is small, as compared with $\sqrt{K_2/\mu}$, its contribution to the slope of VSE is still important. The above difference comes from the contribution of $K_5$ term, which is generally neglected. We find that without $K_5$ and using the Morse potential, one have 
\begin{equation}
\eta = -\frac{a\hbar}{12}\sqrt{\frac{1}{2\mu D_e}}=-\frac{1}{24}\frac{\hbar\omega}{D_e}\sim -0.23\%.
\end{equation}
This value can be found in the appendix of Ref. \cite{Lambert1988vibrational}, which varies from 0.1\% to 0.3\%. When $K_5$ is taken
into account, we have 
\begin{equation} 
\eta = +\frac{7a\hbar}{6}\sqrt{\frac{1}{2\mu D_e}}=\frac{7}{12}\frac{\hbar\omega}{D_e}\sim +3\%.
\end{equation}
In above, we have used $\mu=12\times16/(12+16)\mu_H = 6.857\mu_H$, $D_e =11$ eV,
$\hbar\omega = 0.248$ eV $\sim$ 2000 cm$^{-1}$, and $a=1.54$ \AA, which have been used in Lambert's original manuscript \cite{Lambert1984Stark, herzberg2013molecular}. Thus we see the importance of $K_5$ in the determination of the correct quantum effect. 

Secondly, let us clarify the consequence of the above result. We noticed that in most of the experiments in electrochemistry, the linear Stark effect as a function of potential $\phi$ across the interface has been observed. However, there are still some experiments that reported the weak nonlinear effect \cite{beltramo2005oxidation,zhang2016nonlinear,staffa2017determination,tong2023linear,zwaschka2018microscopic}, which are generally attributed to complicated many-body effect across the interface and which can be understood in some way by the first-principle calculation. When the magnitude of frequency change $\delta \omega$ is much smaller than the vibrational frequency $\omega_0$, that is, $|\delta \omega| \ll \omega_0$, we should only have linear Stark effect; thus the nonlinear effect, in principle, can be observable only when the frequency change is relatively large as compared with the zero-field frequency. From our model, we see that no matter how complicated the interaction is, the atom can only feel the effective field $F \sim 
10^8 - 10^9$ V/m, which is still a weak perturbation (see Eq. \ref{eq-secondorderdw2}) as compared with the atomic unit of electric field of $F_0 \sim 10^{11}$ V/m \cite{shull1959atomic, mcweeny1973natural}. For this reason, assuming that the potential $\phi$ is a complicated yet unknown function of $F$, we should have 
\begin{equation}
{d \omega \over dF} = {\partial \phi \over \partial F} \cdot 
{d \omega \over d\phi}  = \text{constant},
\end{equation}
where $d\omega/d\phi$ is measured directly in experiment. In this way, ${\partial F \over \partial \phi} \propto {d \omega \over d \phi}$. Thus, as a direct logical corollary, the nonlinear Stark effect of $d \omega/d\phi$ should imply that the electric field $F$ across the metal-solution interface is a nonlinear function of potential $\phi$. This result may have been implicitly used in some way in the literature, including Ref. \cite{tong2023linear}, and in this work, we made this point unambiguously and explicitly. This nonlinear relation has already been included in the Gouy-Chapman model across the interface and diffusion layer.
In the future, the above relation should be used as a basic guideline during the understanding of various vibrational spectra in experiments. Furthermore, since the slope $d\omega/dF$ can be determined in theories and $d\omega/d\phi$ can be determined in experiments, we can estimate the relation between $F$ and $\phi$ using this formula. 

Thirdly, we can use this method to estimate the change of binding energy by the electric field. We find that the change in energy is
\begin{equation}
\delta E = -e^* F r_e.
\end{equation}
Here the effect of $\delta r_e \propto F$ is negligible. Using $e^* \sim 0.1e$, we have $\delta E \sim 15$ meV, which is of the order of $10^{-3}D_e$. For this reason, we expect that the electric field will not contribute significantly to the change of binding energy $D_e$. Furthermore, in the above theory, assuming $D_e(F) = D_e(0) + D_e'(0) F$, we expect its contribution to the VSE is of the order of $\mathcal{O}(F^2)$, which is a small quantity as compared with the leading linear term. 

\section{Conclusion and Discussion}
\label{sec-conclusion}

To conclude, this work revisits the VSE and its slope in diatomic molecules by taking the higher-order terms into account using quantum mechanics. We define the quantum effect $\eta$ as the relative difference between quantum and classical VSE and explore its behaviors in various models. For Pt-hydrogen molecules, we find $\eta \sim 8 - 10\%$, while in CO molecules, $\eta \sim 2 -3\%$. Based on this, we revise the Lambert theory, which is about 0.3\% \cite{Lambert1988vibrational, Lambert1996Vibrational}, and find that this small value comes from the neglecting of the higher-order ($K_5$) term. Furthermore, we show that the quantum effect $\eta \sim \chi_e$, where the anharmonic coefficient $\chi_e = \hbar\omega/D_e$. This result provides a new physical interpretation of the anharmonic effect. It also provides a convenient way to determine the role of quantum effect in vibrational spectra, which may be used in future theories. The formulas developed in this work can also be used in the first-principle calculations for a much better understanding of the linear Stark effect and the associated quantum effect. This theory may also be used to understand the large molecules absorbed at the metal surface, which exhibit similar linear VSE. On the experimental side, our results may imply the important role of quantum effect for the VSE in Pt-H, which has a large VSE of about 130 cm$^{-1}$/V \cite{Kunimatsu2006situ, Kunimatsu2007hydrogen}, in which the quantum effect may contribute to about 11 - 13 cm$^{-1}$/V.

This work is not intended to answer the fundamental question of how the quantum effect is related to the functions of the hydrogen evolution reaction and oxygen evolution reaction, which is indeed one of the major starting points of this work. This is a rather complicated issue for the reason that, in the current stage, high-precision experimental results are still limited. In theories, a lot of nonlinear coefficients have been used to interpret the experimental results, however, these nonlinear coefficients can not be directly measured in experiments \cite{wasileski2001metal, Hamada2008Density, koper2011combining}. From our analysis presented in this work, we believe that it is hard to obtain a deterministic relation between the quantum effect, vibrational spectra and their functions of the molecules in evolution reaction. Thus in the future, it should be important on the experimental side to obtain some much more accurate vibrational spectra to fully understand the role of quantum mechanics in this fundamental issue. Our new understanding of the Stark shift may help us to clarify the physics involved in this system. 

\vspace{0.4cm}
\textit{Acknowledgments}: This work was supported by National Natural Science Foundation of China (no. 22172151) and the Innovationa Program for Quantum Science and Technology (Grants No. 2021ZD0301200 and No. 2021ZD0301500).

\appendix           
\section{More details about $\eta$}
\label{sec-appendix}

By definition, the quantum effect $\eta$ is a dimensionless quantity, and it is calculated and given in Eq. \ref{eta-morse} - Eq. \ref{eta-mie}, which can be understood through dimension analysis. In the Morse potential, we expect it is a function of $a$, $\hbar$, $D_e$ and $\mu$. To this end, we can define 
\begin{equation}
\eta \sim D_e^\alpha  \mu^{\beta} a^\gamma \hbar^\delta,
\end{equation}
where the dimensions are $[D_e] = \text{eV}$, $[\mu] = \text{kg}$, $[a] = \text{m}$, and $[\omega] = \text{1/s}$. Using $[\text{kg}] \cdot [\text{(m/s)}^2] = \text{eV}$, we find that the only solution is:
\begin{equation}
\alpha = -{1\over 2}, \quad \beta = -{1\over 2}, \quad \gamma =1, \quad \delta=1.
\end{equation}
Thus we should expect  
\begin{equation}
\eta \sim \frac{a\hbar}{\sqrt{D_e\mu}} \sim {\hbar\omega \over D_e},
\end{equation}
which is used in the main text. For the other two potentials, we can also arrive at the result by replacing $a$ with $1/r_e$, thus we have 
\begin{equation}
\eta \sim \frac{\hbar}{r_e\sqrt{D_e\mu}} \sim {\hbar\omega \over D_e}.
\end{equation}
In all these models, $\chi_e=\hbar\omega/D_e$. 

In the main text, we show that $\eta$ can be different, including its sign, by different types of approximation of the polynomial potentials. Next, we expect the quantum effect from different terms in the three effective models and we find 
\begin{eqnarray}
 \eta &=& \frac{(30K_3^3-26K_2K_3K_4+5K_2^2K_5)\hbar}{K_2^2K_3\sqrt{K_2\mu}},\\
 \eta_2 &=& \frac{(30K_3^3-20K_2K_3K_4)\hbar}{K_2^2K_3\sqrt{K_2\mu}},\\
 \eta_3 &=& \frac{(-6K_2K_3K_4+5K_2^2K_5)\hbar}{K_2^2K_3\sqrt{K_2\mu}},
\end{eqnarray}
in which $\eta_2$ and $\eta_3$ are results from $K_3^{\prime}$ and $K_4^{\prime}$, and $K_3^{\prime}$ and $K_4^{\prime}$, respectively. For the Morse potential, we have 
\begin{equation}
{\eta\over\chi_e} = {7 \over 12}, \quad {\eta_2\over\chi_e} ={5 \over 6}, \quad {\eta_3\over\chi_e} =-{1\over 4},
\end{equation}
which is independent of system parameters. For the LJ potential, we have
\begin{eqnarray}
    {\eta\over\chi_e}&=&\frac{7}{12}+\frac{n-3}{24 n^2},\\
    {\eta_2\over\chi_e}&=&\frac{5}{6}-\frac{5}{6 n^2},\\
    {\eta_3\over\chi_e}&=&-\frac{1}{4}+\frac{n+17}{24 n^2},
\end{eqnarray}
which depend strongly on $n$. For the Mie potential with two parameters, we  have 
\begin{eqnarray}
    {\eta\over\chi_e}&=&[\frac{-9+n(-2+3n)}{12(3+n+An)An^2}\nonumber\\
    &&+\frac{-2+3(5+A)n+7(1+A)n^2}{12n(3+n+An)}],\\ 
    {\eta_2\over\chi_e}&=&\frac{5(-2+An^2)}{6An^2},\\
    {\eta_3\over\chi_e}&=&\frac{n (-A (n-1) (A n+n+6)+n+6)+17}{4An^2(A n+n+3)}.\nonumber\\
\end{eqnarray}
In this expression, $A$ and $n$ are not independent. For example, let the Mie potential and Morse potential have the same second-order harmonic potential, one can obtain $A n^2 = 2 a^2 r_e^2$. With these three expressions, we can determine the role of $K_3$, $K_4$ and $K_5$ for $\eta$ quantitatively. 

One should notice that for the diatomic molecules, as summarized in Table \ref{table-nonH} and \ref{table-Hmoculars}, the parameter $n$ is around 1 to 4. For the Morse and LJ potential, their difference in $\eta/\chi_e$ is determined by $|(n-3)/24n^2|$, ranging from 0 to 1/12, which is much smaller than 7/12. This explains why the fitted lines in Fig. \ref{fig-fig3}(c) and (d) are very close to each other. Furthermore, the sign of $\eta$ and $\eta_{2,3}$ depends strongly on the approximation used in the polynomial potentials, which has not been considered in the previous literature. Finally, while in the large $n$ limits, all $\eta/\chi_e$ approach 7/12 in the three models, which can be regarded as some kind of coincidence, since in this limit the nonlinear terms may become very important, leading to inaccurate approximations in the perturbation theory. This limit can not be reached using the parameters in the two tables \ref{table-nonH} and \ref{table-Hmoculars}. 

\begin{table*}[htp]
	\centering
	\caption{Summary of spectra constants and the associated parameters for the Morse and LJ potential. The second to fifth columns give the experimental results, including spectra constants, $\omega/(2\pi c)$ (in a unit of $\rm{cm}^{-1}$) and $\chi_e\omega/(8\pi c)$ ($\rm{cm}^{-1}$), thermal dissociation energy $D_0$ (eV), equilibrium bond length $r_e$ (\AA). $D_e$ (eV) is the dissociation/binding energy, which is obtained by fitting the experimental results, and $a$ (\AA$^{-1}$) is the parameters used in Morse potential. In addition, the anharmonic coefficient $\chi_e=\hbar\omega/D_e$, 
    and the ratio of quantum effect $\eta$ from the three potentials are presented in the last three columns. This table presents non-hydride diatomic molecules. The parameters are from Ref. \cite{linstrom2001nist} and the $K_5$ of Dunham potential are from Ref. \cite{Noorizadeh2004Recursion}.}
	\begin{tabular}
		{|p{0.05\textwidth}|p{0.07\textwidth}|p{0.09\textwidth}|p{0.06\textwidth}|p{0.06\textwidth}|p{0.08\textwidth}|p{0.08\textwidth}|p{0.08\textwidth}|p{0.08\textwidth}|p{0.08\textwidth}|p{0.08\textwidth}|p{0.08\textwidth}|} \hline
	    &$\omega/(2\pi c)$ & $\chi_e\omega/(8\pi c)$ & $D_0$ & $r_e$ &$D_e$  &$a$ &$n$ &$\hbar\omega/D_e$ &$\eta_{\rm{D}}$ &$\eta_{\rm{Morse}}$ &$\eta_{\rm{LJ}}$ \\ \hline
	LiF &910.34 &7.929 &5.91 &1.563 &5.96 &1.11 &1.74 &0.0185 &0.0123 &0.0108 &0.0105  \\ \hline
        LiCl &643.31 &4.501 &4.84 &2.020 &4.52 &0.90 &1.76 &0.0161 &0.0102 &0.00940 &0.00913   \\ \hline
        LiBr &563.16 &3.53 &4.33 &2.170 &4.36 &0.81 &1.76 &0.0157 &0.00761 &0.00921 &0.00895  \\ \hline
        LiI &498.16 &3.39 &3.54 &2.391 &3.57 &0.79 &1.89 &0.0170 &0.00802 &0.00995 &0.00973  \\ \hline
        NaF &536 &3.4 &5.33 &1.925 &5.36 &0.99 &1.91 &0.0122 &0.00676 &0.00713 &0.00698  \\ \hline
        NaCl &366 &2.05 &4.23 &2.360 &4.25 &0.88 &2.08 &0.0105 &0.00684 &0.00615 &0.00606  \\ \hline
        NaBr &302.1 &1.50 &3.74 &2.502 &3.75 &0.77 &1.94 &0.00986 &0.00677 &0.00575 &0.00564  \\ \hline
        NaI &258 &1.08 &3.00 &2.711 &3.01 &0.74 &2.01 &0.0105 &0.00597 &0.00613 &0.00602  \\ \hline
        KCl &281 &1.30 &4.34 &2.666 &4.35 &0.77 &2.06 &0.00792 &0.00509 &0.00462 &0.00454  \\ \hline
        KBr &213 &0.80 &3.91 &2.820 &3.92 &0.73 &2.06 &0.00668 &0.00396 &0.00389 &0.00383  \\ \hline
        KI &186.53 &0.574 &3.31 &3.047 &3.32 &0.69 &2.13 &0.00692 &0.00413 &0.00403 &0.00398  \\ \hline
        RbF &376 &1.9 &5.00 &2.270 &5.02 &1.42 &3.24 &0.00918 &0.00582 &0.00535 &0.00536  \\ \hline
        RbCl &228 &0.92 &4.34 &2.786 &4.35 &0.93 &2.59 &0.00644 &0.00387 &0.00375 &0.00374   \\ \hline
        RbBr &169.46 &0.463 &3.90 &2.944 &3.91 &0.73 &2.15 &0.00534 &0.00336 &0.00311 &0.00307   \\ \hline
        RbI &138.51 &0.335 &3.30 &3.176 &3.30 &0.72 &2.31 &0.00516 &0.00291 &0.00301 &0.00298   \\ \hline
        CsF &352.56 &1.615 &5.15 &2.345 &5.17 &1.66 &3.91 &0.00837 &0.00456 &0.00488 &0.00490   \\ \hline
        CsCl &214.17 &0.731 &4.58 &2.906 &4.59 &1.07 &3.13 &0.00574 &0.00318 &0.00334 &0.00335   \\ \hline
        CsBr &149.66 &0.374 &4.17 &3.072 &4.17 &0.79 &2.43 &0.00441 &0.00289 &0.00257 &0.00255   \\ \hline
        CsI &119.178 &0.2505 &3.56 &3.315 &3.56 &0.68 &2.26 &0.00412 &0.00248 &0.00240 &0.00238  \\ \hline
	\end{tabular}
 \label{table-nonH}
\end{table*}

\begin{table*}[htp]
	\centering
	\caption{The same as that in Table \ref{table-nonH} for hydride diatomic molecules. The parameters are from Ref. \cite{linstrom2001nist}.}
    \label{table-Hmoculars}
	\begin{tabular}
		{|p{0.05\textwidth}|p{0.07\textwidth}|p{0.09\textwidth}|p{0.06\textwidth}|p{0.06\textwidth}|p{0.08\textwidth}|p{0.08\textwidth}|p{0.08\textwidth}|p{0.08\textwidth}|p{0.08\textwidth}|p{0.08\textwidth}|} \hline
	    &$\omega/(2\pi c)$ & $\chi_e\omega/(8\pi c)$ & $D_0$ & $r_e$ &$D_e$  &$a$ &$n$ &$\hbar\omega/D_e$ &$\eta_{\rm{Morse}}$ &$\eta_{\rm{LJ}}$ \\ \hline
	LiH &1405.65 &23.20 &2.42871 &1.595 &2.50 &1.15 &1.84 &0.0672 &0.0392 &0.0382  \\ \hline
        PtH &2294.68 &46 &3.44 &1.528 &3.58 &1.56 &2.39 &0.0762 &0.0444 &0.0441  \\ \hline
        CuH &1941.26 &37.51 &2.73 &1.462 &2.85 &1.48 &2.17 &0.0811 &0.0473 &0.0467  \\ \hline
        GeH &1833.77 &37 &3.3 &1.588 &3.41 &1.28 &2.03 &0.0639 &0.0372 &0.0366  \\ \hline
        HCl &2990.09 &52.82 &4.4336 &1.274 &4.61 &1.80 &2.30 &0.0774 &0.0451 &0.0447   \\ \hline
        OH &3737.76 &84.89 &4.392 &0.969 &4.62 &2.23 &2.16 &0.0956 &0.0558 &0.0551  \\ \hline
        NH &3282.27 &78.3 &3.07 &1.03 &3.27 &2.32 &2.41 &0.118 &0.0690 &0.0685  \\ \hline
        HBr &2648.98 &45.22 &3.758 &1.414 &3.91 &1.74 &2.46 &0.0810 &0.0472 &0.0469  \\ \hline
        HI &2309.01 &39.64 &3.0541 &1.609 &3.19 &1.68 &2.70 &0.0865 &0.0504 &0.0503  \\ \hline
        HF &4138.32 &89.88 &5.869 &0.916 &6.12 &2.15 &1.97 &0.0800 &0.0467 &0.0458  \\ \hline
        NaH &1172.20 &19.72 &1.88 &1.887 &1.95 &1.09 &2.06 &0.0719 &0.0419 &0.0412  \\ \hline
        KH &983.60 &14.3 &1.86 &2.242 &1.92 &0.92 &2.08 &0.0616 &0.0359 &0.0354  \\ \hline
        CsH &891.00 &12.9 &1.81 &2.493 &1.86 &0.85 &2.12 &0.0575 &0.0335 &0.0330   \\ \hline
        SiH &2041.8 &35.51 &3.55 &1.520 &3.67 &1.38 &2.10 &0.0664 &0.0387 &0.0382  \\ \hline
        PH &2365.2 &44.5 &3.02 &1.422 &3.16 &1.72 &2.45 &0.0890 &0.0519 &0.051   \\ \hline
        SH &2711.6 &59.9 &3.55 &1.340 &3.71 &1.81 &2.42 &0.0863 &0.0503 &0.0500   \\ \hline
        BH &2366.9 &49.40 &3.42 &1.232 &3.56 &1.61 &1.99 &0.0788 &0.0459 &0.0451   \\ \hline
        AlH &1682.56 &29.09 &3.06 &1.647 &3.16 &1.23 &2.02 &0.0636 &0.0371 &0.0364   \\ \hline
        GaH &1604.52 &28.77 &2.84 &1.663 &2.94 &1.21 &2.02 &0.0652 &0.0380 &0.0374   \\ \hline
        InH &1476.04 &25.61 &2.48 &1.838 &2.57 &1.19 &2.20 &0.0686 &0.0400 &0.0395  \\ \hline
        ZnH &1607.6 &55.14 &0.851 &1.594 &0.95 &2.06 &3.29 &0.194 &0.113 &0.113  \\ \hline
        AgH &1759.9 &34.06 &2.28 &1.618 &2.39 &1.47 &2.38 &0.0877 &0.0511 &0.0508  \\ \hline
        AuH &2305.01 &43.12 &3.32 &1.523 &3.46 &1.60 &2.44 &0.0794 &0.0463 &0.0460  \\ \hline
        NiH &1926.6 &38 &3.07 &1.475 &3.19 &1.39 &2.05 &0.0719 &0.0419 &0.0412  \\ \hline
        MnH &1548.0 &28.8 &2.5 &1.731 &2.59 &1.24 &2.15 &0.0711 &0.0415 &0.0409  \\ \hline
        BeH &2060.78 &36.31 &2.034 &1.342 &2.15 &1.82 &2.44 &0.114 &0.0666 &0.0661  \\ \hline
        MgH &1495.20 &31.89 &1.34 &1.729 &1.43 &1.61 &2.78 &0.123 &0.0722 &0.0720  \\ \hline
        CaH &1298.34 &19.10 &1.70 &2.002 &1.78 &1.27 &2.54 &0.0877 &0.0511 &0.0509  \\ \hline
        SrH &1206.2 &17.0 &1.66 &2.145 &1.73 &1.19 &2.57 &0.0837 &0.0488 &0.0486  \\ \hline
	\end{tabular}
\end{table*}

\bibliography{ref.bib}

\end{document}